\documentclass[journal]{IEEEtran} 
\usepackage[utf8]{inputenc}
\usepackage{amsmath}
\usepackage{amssymb}
\usepackage{mathtools}
\usepackage{graphicx}
\usepackage{optidef}
\usepackage{amsfonts}
\usepackage{algorithm}
\usepackage{algorithmic}
\usepackage{caption}
\usepackage{subcaption}
\usepackage{amsthm}

\usepackage{xcolor}
\usepackage{breqn}
\usepackage{cuted}
\usepackage{tabularx}
\usepackage{multicol}
\usepackage{colortbl}

\ifCLASSINFOpdf
\else
\fi
\DeclareUnicodeCharacter{2212}{-}

\begin{document}
%
\title{Optimizing URLLC in Open RAN: A Deep Reinforcement Learning-Based Trade-off Analysis}
%
%
%

\author{Rana~Muhammad~Sohaib,~\IEEEmembership{Member,~IEEE,}
        Syed~Tariq~Shah,~\IEEEmembership{Member,~IEEE,}
    Muhammad~Ali~Jamshed,~\IEEEmembership{Senior~Member,~IEEE,}
    Oluwakayode~Onireti,~\IEEEmembership{Senior~Member,~IEEE,}
    Poonam~Yadav,~\IEEEmembership{Senior~Member,~IEEE}
\thanks{Rana M Sohaib and Poonam Yadav are with the University of York, e-mail: \{rana.sohaib, poonam.yadav\}@york.ac.uk. M. A. Jamshed and Oluwakayode Onireti are with the University of Glasgow, UK, e-mail: \{muhammadali.jamshed, Oluwakayode.Onireti\}@glasgow.ac.uk). S. S. Tariq, is with the University of Essex, e-mail: Syed.Shah@essex.ac.uk}
}




\maketitle

\begin{abstract}
The advent of Ultra-Reliable Low Latency Communication (URLLC) along with the emergence of Open RAN (ORAN) architectures presents unprecedented challenges and opportunities in Radio Resource Management (RRM) for next-generation communication systems. This paper presents a comprehensive trade-off analysis of Deep Reinforcement Learning (DRL) approaches designed to enhance URLLC performance within ORAN's flexible and dynamic framework. By investigating various DRL strategies for optimizing RRM parameters, we explore the intricate balance between reliability, latency, and the newfound adaptability afforded by the ORAN principles. Through extensive simulation results, our study compares the efficacy of different DRL models in achieving URLLC objectives in an ORAN context, highlighting the potential of DRL to navigate the complexities introduced by ORAN. 
The proposed study provides valuable information on the practical implementation of DRL-based RRM solutions in ORAN-enabled wireless networks. It sheds light on the benefits and challenges of integrating DRL and ORAN for URLLC enhancements. Our findings demonstrate that the proposed twin-delayed deep-deterministic policy gradient (TD3) integrated with Thompson Sampling (TS) achieves reliability levels above 99\% in more than 80\% of instances, outperforming baseline DRL methods in maintaining stringent URLLC reliability requirements, offering a roadmap for future research to pursue efficient, reliable, and flexible communication systems.

\end{abstract}

\begin{IEEEkeywords}
URLLC, ORAN, DRL, RRM, power allocation.
\end{IEEEkeywords}

%
\IEEEpeerreviewmaketitle

\vspace{-0.3cm}
\section{Introduction}

\IEEEPARstart{O}{pen} radio access network (O-RAN) is a revolutionary approach to designing and deploying mobile networks, aiming to disaggregate traditional, proprietary network elements and foster interoperability among different vendors' hardware and software components. In a conventional RAN, the hardware and software are tightly integrated, often supplied by a single vendor. O-RAN, however, promotes a more open and flexible ecosystem by separating the radio network functions into standardised, interoperable components. This separation allows network operators to mix and match components from different vendors, fostering competition, reducing dependency on a single supplier, and potentially lowering costs.
In an Open RAN architecture, the radio functions are divided into three main components: the Radio Unit (RU), the Distributed Unit (DU), and the Centralized Unit (CU). The RU contains radio transceivers and antennas, the DU handles baseband processing, and the CU manages higher-layer functions. Open interfaces and standardised protocols facilitate interoperability and allow operators to deploy best-of-breed solutions for each network element. This approach is particularly appealing for promoting innovation, enabling new market entrants, and fostering a more dynamic and competitive ecosystem within the telecommunications industry. While O-RAN promises increased flexibility and cost efficiency, it also presents challenges, including the need for robust standardisation, addressing interoperability issues, and ensuring that mobile networks' performance and security requirements, especially in terms of latency and reliability, are met in diverse deployment scenarios. In response to these challenges, the RAN Intelligent Controller (RIC) has emerged as a transformative technology within the 5G ecosystem. 
The RIC facilitates a modular and open architecture to manage resources efficiently and enhance the overall user experience by leveraging advanced algorithms, machine learning, and data analytics, promoting interoperability and innovation by allowing third-party applications to interact with the RAN through standardised interfaces. 

The landscape of wireless communication is transforming with the proliferation of Ultra-Reliable Low Latency Communication (URLLC). One of the key features of 5G is the support for Ultra-Reliable and Low-Latency Communications (URLLC), which is expected to revolutionise industries such as autonomous driving, industrial automation, and healthcare by enabling real-time and reliable communication \cite{b1, b2}. However, the stringent Quality of Service (QoS) requirements of URLLC, such as high reliability and low latency, pose significant challenges to the design and management of radio resources \cite{b3}. URLLC caters to applications that require stringent reliability and low latency, such as autonomous vehicles, industrial automation, and mission-critical healthcare systems \cite{b4}. 
The demand for URLLC is driven by the surge in applications requiring instantaneous and error-free communication. For instance, vehicles must exchange real-time information in autonomous driving scenarios to make split-second decisions, demanding ultra-reliable and low-latency communication. Similarly, in industrial automation, where machines collaborate in real time, any communication delay or failure can have severe consequences. Recognising the transformative potential of URLLC applications, the industry is actively seeking solutions to overcome the challenges posed by these stringent requirements.
As URLLC applications become more prevalent, communication systems must evolve to meet the stringent requirements imposed by these use cases. 
In this context, Radio Resource Management (RRM) is pivotal in ensuring efficient spectrum utilisation, interference management, power allocation, and overall network performance \cite{b5,b6}. Achieving the delicate balance between ultra-reliability and low latency is a complex challenge that demands innovative solutions. Traditional RRM techniques, however, may need to be revised to meet the stringent requirements of URLLC due to their inability to adapt to dynamic network conditions and user requirements \cite{b8}. Achieving reliability levels on the order of 99.99\%.
URLLC necessitates reevaluating the trade-offs between conflicting objectives, such as optimising spectral efficiency while minimising latency. The coexistence of different types of traffic, each with very stringent but completely different requirements, presents a novel and very relevant research issue. Moreover, the dynamic and unpredictable nature of URLLC traffic patterns requires adaptive and intelligent RRM solutions that can respond in real-time to changing conditions \cite{b9}. 

Deep Reinforcement Learning (DRL), a subset of machine learning, has emerged as a promising approach for addressing optimisation problems in wireless communication \cite{b10}. With its ability to learn optimal strategies through interaction with the environment, DRL presents a promising avenue for addressing the challenges to achieve URLLC. By leveraging DRL techniques, it becomes possible to develop intelligent systems that adapt to the unique requirements of URLLC applications, optimising RRM parameters in a manner that balances reliability and latency. It has been applied to various aspects of URLLC, including radio resource allocation and power allocation \cite{b11}, data management on top of scheduled eMBB traffic \cite{b12}, and resource allocation for diverse URLLC \cite{b9, b13}. Several studies have proposed DRL-based approaches for URLLC \cite{b9, b13}.  
Despite these advancements, there are still open challenges and future research directions in applying DRL techniques in RRM for URLLC applications. While the literature on DRL applications in communication systems is expanding rapidly, a noticeable gap exists in the specific domain of URLLC. Notably, more comprehensive studies that systematically explore the trade-offs inherent in various DRL approaches for optimising RRM parameters in URLLC scenarios need to be conducted. 
The need for a comprehensive trade-off analysis becomes increasingly evident as URLLC applications become more prevalent, necessitating adaptive and intelligent RRM solutions.
Existing research often focuses on isolated aspects of DRL or URLLC. However, a holistic examination of the trade-offs between different DRL strategies for RRM optimisation in URLLC applications is notably required. Addressing this gap is crucial for advancing state-of-the-art URLLC communication systems, as it provides a foundation for understanding the strengths and weaknesses of various DRL approaches. This knowledge is essential for guiding the design and deployment of communication systems that can meet the stringent requirements of URLLC applications without compromising reliability or latency.

This paper aims to enhance URLLC performance by conducting a trade-off analysis of DRL approaches for radio resource management. It comprehensively reviews the most widely used DRL algorithms to address resource allocation and power allocation problems, including the value- and policy-based algorithms in next-generation wireless networks. Each algorithm's advantages, limitations, and use cases are thoroughly discussed. Furthermore, this study aims to fill the void in the literature by offering a comprehensive exploration of the nuanced relationships between DRL strategies and key URLLC performance metrics. Finally, the paper highlights critical open challenges and provides insights into several future research directions.
\vspace{-0.30cm}
\section{DRL-powered RRM Approaches}
In literature, a spectrum of DRL approaches is considered for optimising RRM parameters in URLLC scenarios, including deep deterministic policy gradient (DDPG), policy gradient actor-critic (PGAC), double deep Q-networks (DDQN), and deep Q-networks (DQN). 
Each DRL approach is tailored to address the specific challenges of RRM optimisation, incorporating state-of-the-art techniques in the reinforcement learning domain. In this section, we highlight the challenges associated with these DRL approaches.
\vspace{-0.4cm}
\subsection{Deep deterministic policy gradient (DDPG)}
DDPG employs an actor-critic architecture. The actor-network learns a policy that maps state observations to actions, while the critic network evaluates the value of the chosen actions. DDPG uses target networks for both the actor and critic to enhance stability during training. These target networks slowly track the learned networks' parameters. It enables end-to-end learning, allowing the algorithm to directly optimise the RRM parameters based on the observed performance metrics and rewards \cite{b11}. Unlike traditional reinforcement learning algorithms that use stochastic policies, DDPG uses a deterministic policy. This policy selection means the actor network outputs a deterministic action for a given state.
DDPG is well-suited for problems with continuous action spaces, making it applicable to fine-grained RRM parameter adjustments.
\subsubsection*{Challenges}
Training DDPG may require a large number of samples, which could be a challenge in scenarios where collecting data is costly or time-consuming \cite{b18}. DDPG relies on exploration strategies such as adding noise to actions, which may not always be effective, especially in complex environments with sparse rewards.
\vspace{-0.5cm}
\subsection{Policy Gradient Actor-Critic (PGAC)}
Policy Gradient methods, particularly the Actor-Critic architecture, have shown promise in optimising sequential decision-making processes in scheduling URLLC traffic \cite{b9}. PGAC involves two key components: the actor, which learns a policy to map states to actions, and the critic, which evaluates the policy's performance and guides the learning process.
PGAC is implemented with a neural network architecture where the actor and critic components are jointly trained. The actor network learned a policy that influenced RRM decisions, while the critic assessed the quality of these decisions. Using a continuous action space in PGAC facilitated nuanced adjustments to RRM parameters.
\subsubsection*{Challenges}
One of the primary challenges associated with PGAC is the issue of high variance in gradient estimates, often leading to slow convergence and suboptimal policies. PGACL methods inherently explore the action space by sampling actions according to the policy distribution. However, ensuring sufficient exploration can be challenging, especially in wireless communication environments with complex action spaces.
Despite its merits, PGAC may suffer from sample inefficiency, requiring considerable data for effective learning. Additionally, the sensitivity to hyperparameter tuning poses a challenge in achieving optimal performance across diverse URLLC scenarios \cite{b13}.
\vspace{-0.3cm}
\subsection{Double Deep Q-Networks (DDQN)}
DDQN is an extension of the traditional Q-learning algorithm that mitigates overestimation biases by employing two separate networks for action selection and value estimation. This dual-network architecture aims to stabilise the learning process and improve the accuracy of action value predictions. 
DDQN exhibited improved training stability compared to traditional Q-learning approaches, addressing concerns related to overestimating Q-values and enhancing the robustness of the learned policies.
Using target networks in DDQN mitigated the issues associated with the moving target problem, resulting in more accurate Q-value estimates.
DDQN excels in scenarios with discrete action spaces, making it suitable for RRM parameter optimisation tasks that involve discrete decision-making. 
\subsubsection*{Challenges}
Despite its benefits, DDQN may face challenges in efficiently handling continuous action spaces, which are prevalent in specific URLLC optimisation problems \cite{b18}.
\subsection{Deep Q-Networks (DQN)}
DQN serves as a foundational algorithm in reinforcement learning, offering simplicity and ease of implementation.
Its discrete action space compatibility suits certain RRM scenarios where discrete decisions are prevalent.
\subsubsection*{Challenges}
The discretisation of action spaces could limit the precision of the learned policies, especially in URLLC scenarios where fine-grained adjustments are crucial.
DQN's reliance on a single Q-network could lead to overestimation bias, impacting the accuracy of learned Q-values and potentially affecting the model's overall performance \cite{b15}.

\section{DRL in URLLC: Concepts and Framework}
\begin{figure}[t]
     \centering
\includegraphics[width=0.45\textwidth]{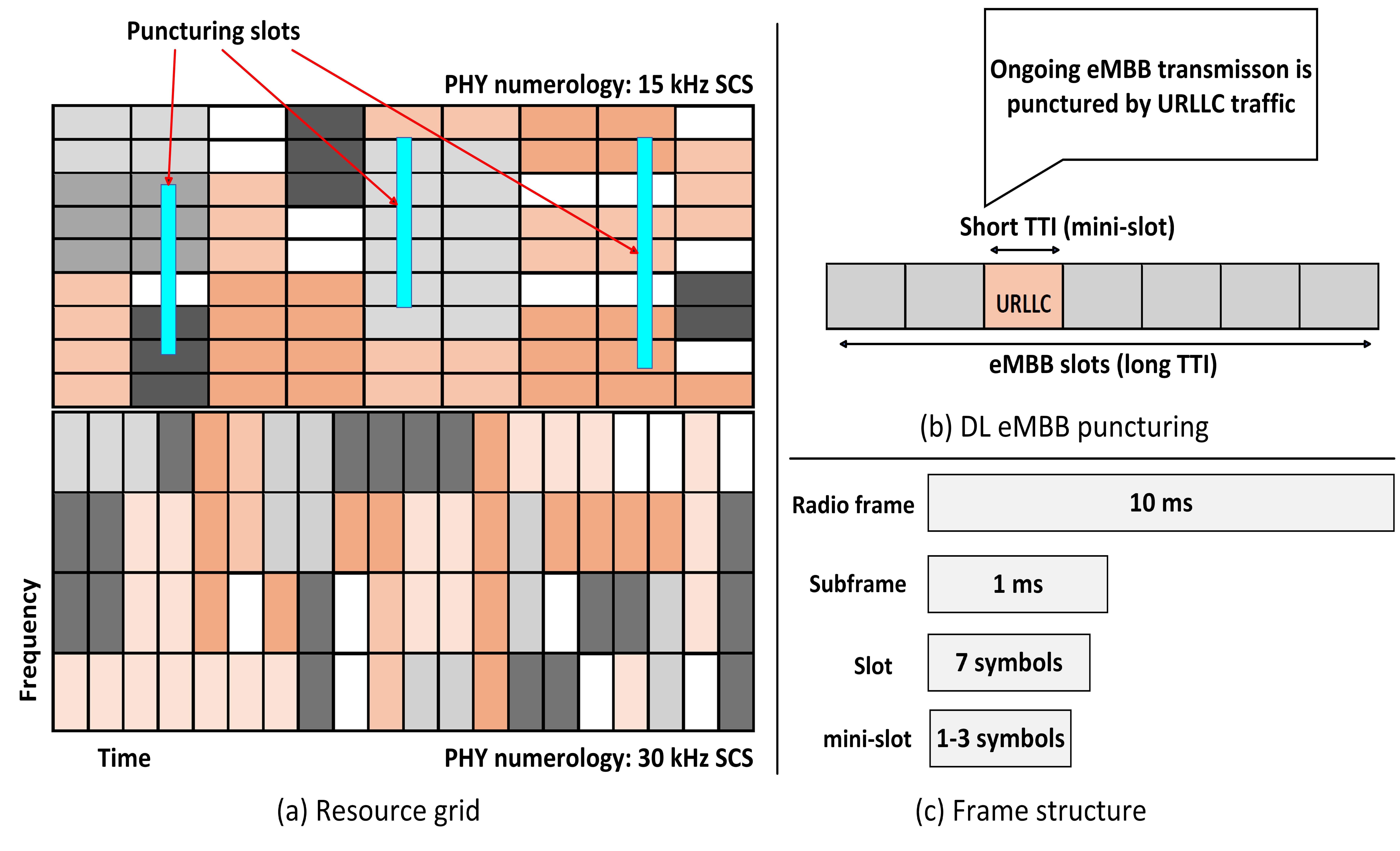}
\caption{Illustration of the puncturing strategy: a) resource grid; b) DL eMBB puncturing; c) frame structure.}
\label{fig:1}
\end{figure}

Due to the critical nature of URLLC traffic, which cannot be delayed during ongoing eMBB communication, we adopt a puncturing strategy. The scheduling approach involves puncturing the eMBB slots to accommodate the transmission of URLLC traffic during short Transmission Time Intervals (TTIs). URLLC services are scheduled at a short TTI duration of 0.5 ms, emphasising low-latency communication. When both eMBB and URLLC requests coexist, eMBB slots are punctured to facilitate the transmission of URLLC traffic without delay.
The puncturing strategy ensures that the stringent latency requirements of URLLC traffic are met, even during ongoing eMBB communication. Fig. \ref{fig:1} illustrates the puncturing strategy where the eMBB slots are punctured by URLLC traffic. 
The integration of DRL into URLLC presents a promising avenue for addressing the unique challenges associated with optimising RRM parameters. 
DRL, a subset of machine learning, is characterised by its ability to learn optimal decision-making policies through interaction with an environment. DRL has gained immense attention in communication systems for its potential to adapt and optimise system parameters based on real-time feedback. Within the URLLC landscape, where the demand for reliability and low latency is paramount, DRL emerges as a potent tool for crafting adaptive and intelligent RRM strategies.
The core concept of DRL involves an agent interacting with an environment, receiving feedback in the form of rewards or penalties, and adjusting its actions to maximise cumulative rewards over time. In the URLLC context, the environment encompasses communication channels' dynamic and unpredictable nature, varying traffic patterns, and stringent reliability and latency requirements.
To address the challenges URLLC poses, DRL must be customised to the unique characteristics of these applications. The framework involves defining states, actions, and rewards that capture the intricacies of URLLC performance metrics. States encapsulate relevant information about the communication system, actions represent the decisions made by the agent, and rewards reflect the system's performance in meeting URLLC objectives. For instance, in the URLLC domain, states may include channel conditions, traffic load, and historical performance metrics. Actions could encompass resource allocation, scheduling, and interference management decisions. Rewards may be derived from achieving low latency, ensuring high reliability, or optimising spectral efficiency.
\begin{figure*}[t]
     \centering
\includegraphics[width=\textwidth,height=0.6\textheight,keepaspectratio]{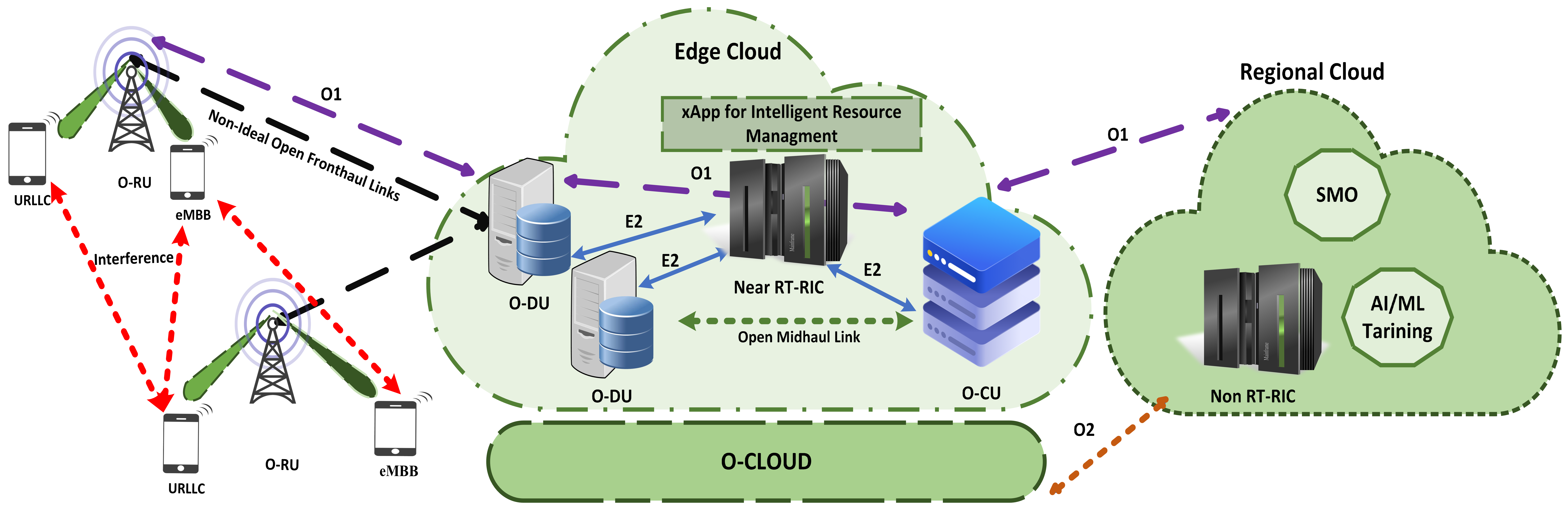}
\caption{Illustration of the system model of use case.}
\label{fig:sys}
\end{figure*}
\vspace{-0.3cm}
\subsection{Conceptual Framework for DRL-based RRM in URLLC}
Fig. \ref{fig:sys} illustrates the system model of the use case. 
The conceptual framework for applying DRL to RRM in URLLC involves a closed-loop process \cite{b10}. The agent observes the current state of the communication system, selects actions based on learned policies, interacts with the environment, and receives feedback as a reward. 
These states serve as the foundation for decision-making by the DRL agent. In the context of URLLC, states may encompass variables such as channel conditions, traffic load, historical performance metrics, and other relevant parameters that influence the reliability and latency of communication links. Including both instantaneous and historical information allows the DRL agent to make informed decisions considering the dynamic nature of URLLC traffic. The flexibility of the action space enables the DRL agent to adapt its decision-making process to diverse and evolving scenarios. The granularity of the action space is crucial. Fine-tuning actions allow the agent to respond precisely to the specific demands of URLLC applications, ensuring a balance between reliability and latency. For example, priority scheduling or dynamic resource allocation decisions can significantly impact URLLC performance.
This iterative process enables the DRL-based RRM system to continually adapt and optimise its strategies in response to evolving network conditions and URLLC requirements.
A key consideration in the framework is designing an appropriate reward structure that aligns with URLLC objectives. Balancing the trade-offs between reliability and latency becomes central to formulating a reward function that guides the DRL agent toward achieving desired URLLC performance metrics. 
\vspace{-0.3cm}
\subsection{Role of DRL in Optimising RRM Parameters}
Within the DRL-based RRM framework for URLLC, specific attention is given to optimising crucial parameters. These may include, but are not limited to, scheduling policies, resource allocation strategies, and interference mitigation techniques. DRL's ability to adapt and learn from dynamic environments positions it as a powerful tool for crafting policies that balance these parameters to meet the stringent demands of URLLC applications. 
In DRL or decision-making problems, an $\epsilon$-greedy approach is a strategy for selecting actions. The parameter $\epsilon$ represents the probability of choosing a random (exploration) action instead of the one that is currently believed to be the best (exploitation). When $\epsilon$ is small, the agent primarily exploits the current best-known action; when $\epsilon$ is large, the agent explores more by randomly selecting actions. While exploiting the current best-known action might be effective in many situations, it may not always lead to the globally optimal solution. In some cases, the agent may get stuck in a sub-optimal solution, as it tends to repeatedly choose the action with the highest estimated reward, possibly missing out on better alternatives that it has yet to explore \cite {b16}. 

Existing research has primarily concentrated on traditional exploration-exploitation techniques, notably $\epsilon$-greedy algorithms, to address the challenges of radio resource allocation, HO decision-making, and related tasks. The absence of comprehensive studies on other exploration-exploitation techniques, such as Thompson sampling (TS), leaves a critical void in understanding its applicability and efficacy in radio resource management scenarios.  
\vspace{-0.2cm}
\section{The Proposed Methodology}
\subsection{Thompson Sampling (TS) for URLLC}
In \cite{b13}, we show the efficient, intelligent resource scheduling of URLLC users by implementing the TS. 
TS, a robust Bayesian algorithm, presents a promising solution for optimising resource allocation while ensuring the stringent QoS demands of these service categories are met. 
In URLLC, TS proves valuable in scenarios where uncertainties in the communication environment are prevalent. For instance, in the allocation of resources, TS dynamically adapts to changes in channel conditions and traffic patterns by updating its probability distribution, thereby addressing the challenges of URLLC's dynamic and unpredictable nature. 
Using Bayesian inference, TS begins by modelling the uncertainty associated with each action's true underlying reward distribution. It maintains a distribution (often a Beta distribution) for each action, representing the agent's belief about the likelihood of different reward values. During each decision-making iteration, TS samples from these distributions for each action. The action associated
with the highest sample is then chosen for execution. This process reflects a probabilistic approach that naturally balances exploration (sampling from uncertain distributions) and exploitation (choosing the best current option). 
When employing TS, the DRL agent gains the ability to gather knowledge and adjust to these shifting conditions persistently. It can modify its decisions on distributing resources by relying on the most recent data available.
This adaptability is crucial in meeting the stringent requirements of URLLC applications, where network conditions and traffic demands can vary dynamically.
\vspace{-0.3cm}
\subsection{Twin-delayed deep deterministic policy gradient (TD3)}
The TD3 algorithm is an extension of the Deep Deterministic Policy Gradient (DDPG) algorithm and addresses its overestimation bias, which can lead to poor policy performance. TD3 introduces three critical improvements: clipped double-Q learning, delayed policy updates, and target policy smoothing, which collectively help to stabilise the training process and improve the performance of the learned policies \cite{b18}.
Incorporating twin critics in TD3 contributes to increased stability during the learning process. This robustness is essential in dynamic URLLC environments where the wireless channel conditions and interference patterns evolve continuously. The dual critics provide more accurate Q-value estimates, mitigating the risk of overestimation and enhancing the reliability of learned policies.
TD3 demonstrated enhanced sample efficiency compared to traditional policy gradient methods. This characteristic is particularly advantageous in URLLC scenarios where rapid adaptation to changing conditions is crucial. 
\begin{figure}[t]
     \centering
\includegraphics[width=0.5\textwidth]{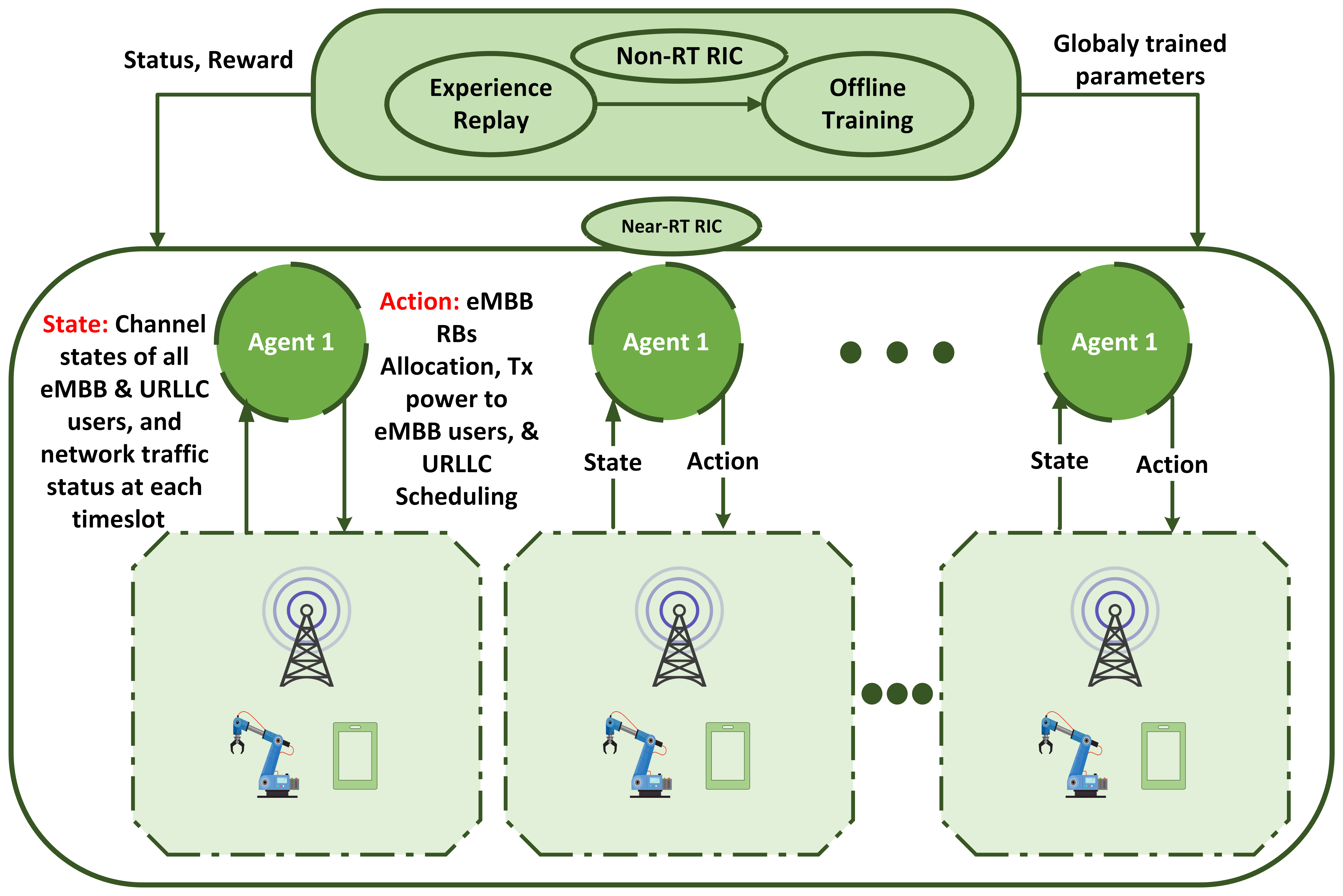}
\caption{Distributed DRL for RRM.}
\label{fig:2}
\end{figure}
\vspace{-0.4cm}
\subsection{Multi-Agent TS-based TD3 Approach}
To address the challenges and leverage the advantages identified in the DRL approaches, we propose a novel methodology that integrates the strengths of TD3 with the exploration strategy of TS. 
We integrate TS into the TD3 framework by incorporating Bayesian techniques to model uncertainties in the RRM decision space. The TS-based TD3 approach maintains the twin critics for stability and leverages TS to guide policy updates, injecting a probabilistic element into the decision-making process. By considering uncertainties in decision-making, the model can respond more effectively to variations in URLLC traffic patterns and wireless channel conditions. The training process involves initialising the TD3 model with TS enhancements and updating policies based on sampled values from the posterior distribution. The training iterations aim to refine the model's understanding of the URLLC environment and improve the robustness of the learned policies.
Fig. \ref{fig:2} refers to the distributed framework, where every RU serves as a DRL agent at near-RT-RIC. Training is implemented offline at a centralised server in non-RT-RIC, where information is gathered from all regional DRL agents. Each agent makes independent decisions, which enables all DRL agents to improve the learning process and convergence rate. The centralised server shares the trained parameters with all DRL agents.  
The global model undergoes training to optimise and maximise a predefined global reward function, which is defined as the achievable eMBB rate and URLLC reliability.
This means that during the training process, the model is continually adjusted to improve its performance based on the specified objective in the global reward function. Every DRL agent decides the optimal resource allocation policy by learning from the trained model according to the observed local environment. The model learns and refines its parameters to enhance its ability to achieve the optimal policy as dictated by the global reward function.  
\begin{table}[b]
\caption{Simulation Parameters} 
\centering
\label{tab:1}
\begin{tabular}{|p{2.75cm}| p{5cm}|}
    \rowcolor{black!10}
\hline
\textbf{Parameters} & \textbf{Values} \\ 
\hline
 Service Area & 4 BSs, each BS covers 200 $m$^2, full-buffer traffic \\ \hline
 System bandwidth & 20 MHz \\ \hline
URLLC packets length & 32 Bytes  \\ \hline
RB Bandwidth & 180 kHz \\ \hline
Transmit power & 38 dBm \\ \hline
Pathloss Model & 120.8 + 37.5 log10(d) \\ \hline
PGAC   & $\delta_a=10^{-5}$, $\delta_c=10^{-3}$
\\ \hline
DDPG  & $\delta_a=10^{-4}$, $\delta_c=10^{-3}$ \\ \hline
TD3  & $\delta_a=10^{-5}$, $\delta_c=10^{-3}$ \\ \hline
\end{tabular}
\end{table}

\vspace{-0.1cm}
\section{DRL-Powered RRM: A Case Study}
URLLC requires highly reliable and low-latency communication, which is essential for applications such as autonomous driving, industrial automation, and remote surgery.
In such environments, the wireless network conditions can be very volatile, and managing resources like power, spectrum, and interference becomes critical.
In this section, we present how to intelligently optimise the radio resources for URLLC incoming traffic. 
Four BSs are configured to form a wireless network, catering to various communication needs within the coverage area of 200 $m^2$. We consider two types of downlink requests: eMBB and URLLC. 
The heterogeneity of the network, combined with the varying nature of eMBB and URLLC requests, provides a realistic and challenging environment for evaluating the DRL approaches. We present the simulation parameters in Table \ref{tab:1}. The proposed and other DRL approaches are trained using different communication configurations, e.g., varying URLLC arrival rates. 
The optimiser chosen for all DRL approaches is the Adam algorithm.
\begin{figure}[t]
     \centering
\includegraphics[width=0.45\textwidth]{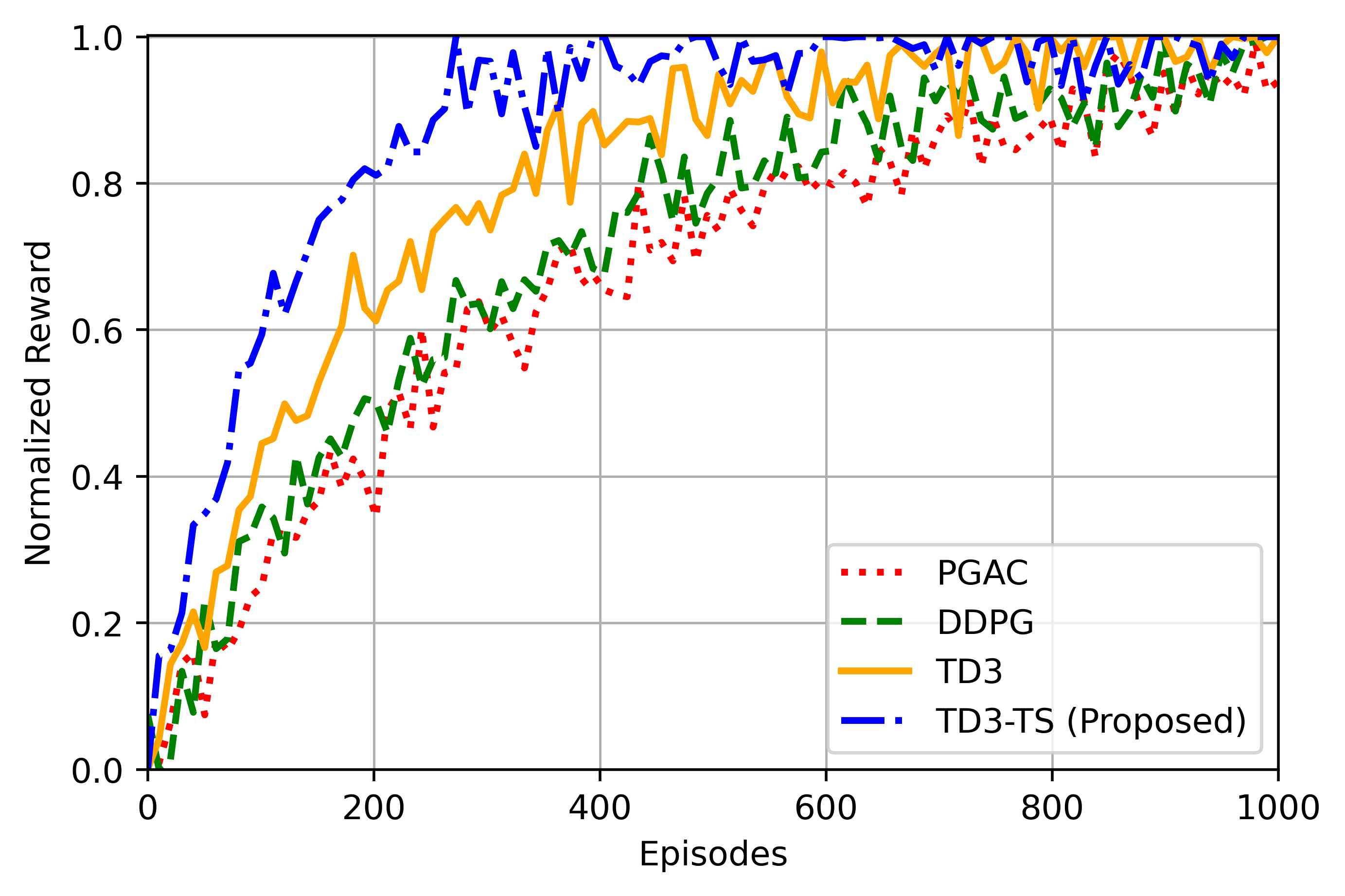}
\caption{Learning Analysis}
\label{fig:5}
\end{figure}
In Fig. \ref{fig:5}, it can be observed that TD3-TS achieves optimal performance earlier and maintains it more consistently, suggesting that it efficiently learns to allocate resources under stringent URLLC constraints, outperforming other DRL methods that show slower or more fluctuating learning behavior. This superiority comes from its ability to balance exploration and exploitation more effectively in dynamic environments, ensuring better reliability and low-latency communication.
In Fig. 5, we investigate the URLLC reliability obtained by different DRL schemes and compare the performance with the proposed approach by plotting the CCDF of outage probability. 
Fig. \ref{fig:fig4} demonstrates that the proposed approach based on TS maintains the outage probability within the tolerable threshold in over 99\% of instances. TS enables the TD-3 agent to incorporate adaptive exploration by sampling from the posterior distribution over the model parameters. This helps it to effectively explore the action space, especially in scenarios where the environment dynamics are uncertain or changing. However, it can be observed from Fig. \ref{fig:fig3}  that as URLLC traffic rates increase, challenges to URLLC reliability may arise. Despite these challenges, our proposed method demonstrates a remarkable ability to maintain reliability in transmitting URLLC packets compared to other DRL approaches.
The inclusion of TS in TD3 helps the agent explore the action space more informedly, leading to more optimal decisions that result in a lower outage probability for URLLC. This approach can be especially beneficial in environments with high uncertainty or where the reward distribution is non-stationary, which might be the case in URLLC scenarios. TD3-TS’s performance, as visualised in the provided CCDF plot, suggests it is more adept at handling the complex and time-sensitive trade-offs required in URLLC scenarios.
\begin{figure}[t]
     \centering
     \begin{subfigure}[b]{0.45\textwidth}
         \includegraphics[width=\textwidth]{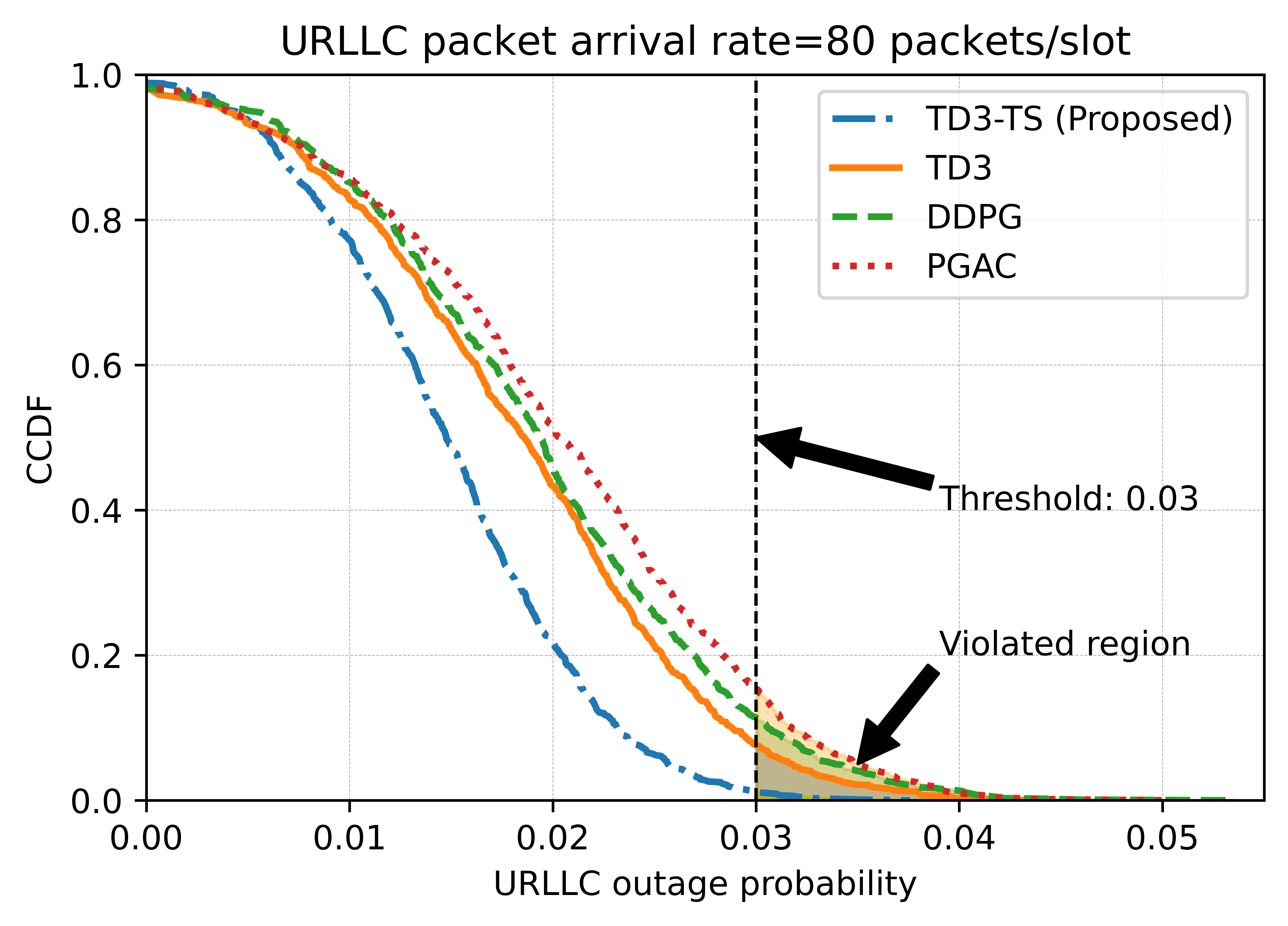}
         \caption{CCDF of URLLC outage probability when $\phi=80$ packets/slot}
         \label{fig:fig4}
     \end{subfigure}
     \hfill
     \begin{subfigure}[b]{0.45\textwidth}
         \includegraphics[width=\textwidth]{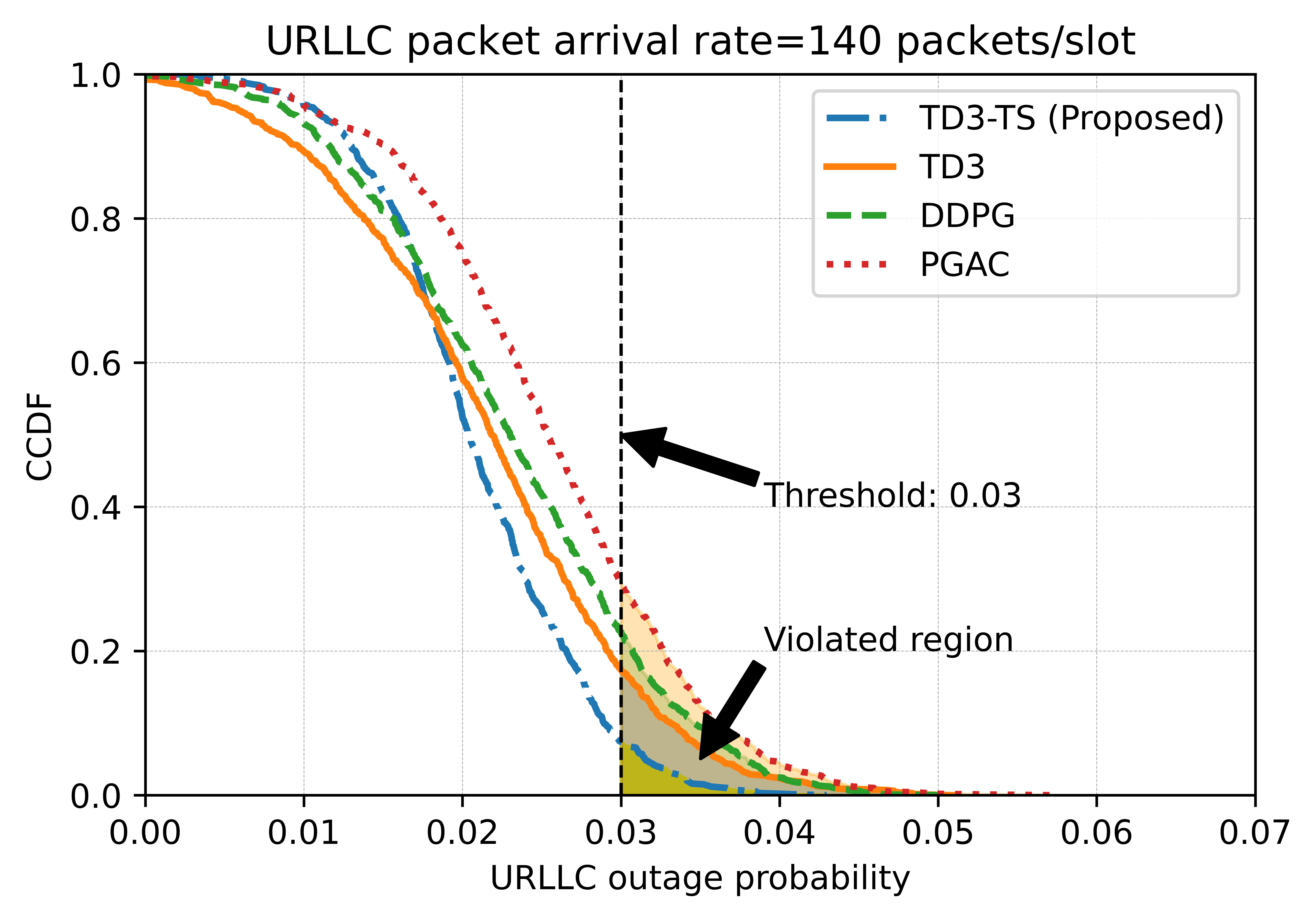}
         \caption{CCDF of URLLC outage probability when $\phi=140$ packets/slot}
         \label{fig:fig3}
     \end{subfigure}
     \caption{CCDF of URLLC reliability for different $\phi$.}
\end{figure} 
\vspace{-0.4cm}
\subsection{Trade-off Analysis}
\subsubsection*{Exploration vs. Exploitation} 
TD3-TS provides a more sophisticated exploration strategy, which could lead to the discovery of more reliable communication strategies in the long term. However, the benefit of improved exploration comes with the trade-off of potentially slower initial learning progress as the algorithm takes time to explore various options before exploiting the best strategy.
Traditional TD3 and DDPG may exploit known-good strategies sooner, but if they become stuck in local optima, this could lead to sub-optimal long-term performance.
\subsubsection*{Complexity vs. Performance}
The integration of TS into TD3 increases the algorithmic complexity, which might require more computational resources or sophisticated training procedures. The improved performance in URLLC scenarios can justify this complexity, but it may only be suitable for some applications, especially those with constrained computational resources.
Simpler algorithms like DDPG might be easier to implement and require less computational power, making them more appropriate for systems with limited processing capabilities despite possibly lower performance.
\subsubsection*{Stability vs. Responsiveness}
TD3-TS's twin Q-networks and delayed policy updates contribute to the learning process's stability. This stability is crucial for URLLC, where erratic behaviour could lead to unacceptable packet loss or delay levels. However, these features might also make the algorithm less responsive to sudden environmental changes.
PGAC and DDPG may adapt more quickly to environmental changes due to their potentially more responsive update mechanisms, but with the risk of higher volatility in performance.
\subsubsection*{Overhead vs. Scalability}
TS requires maintaining and updating a probability distribution over actions, which introduces additional overhead. This might affect the scalability of TD3-TS to very large or complex networks.
Algorithms without such requirements may scale more easily but might not provide the same level of performance in terms of outage probability, as seen in the provided plot.
\section{Conclusion}
This study presented an in-depth analysis of reinforcement learning algorithms' performance in managing URLLC's stringent requirements. This work is significant in its contribution to the advancement of 5G and beyond wireless networks, where URLLC is pivotal for mission-critical applications. These applications demand high reliability and the capability to adapt swiftly to dynamic network conditions.
Our findings highlight the effectiveness of integrating advanced statistical methods such as TS into machine learning algorithms, which can significantly enhance their effectiveness in highly demanding URLLC applications. The TD3-TS approach, with its enhanced learning capabilities and strategic action selection process, paves the way for future explorations into intelligent and autonomous network management solutions. As we move forward, the continuous improvement of such algorithms and their adaptation to the evolving landscape of URLLC will remain a cornerstone of technological progress in the era of interconnected systems.

\ifCLASSOPTIONcaptionsoff
  \newpage
\fi

\vspace{-0.12cm}
\bibliographystyle{IEEEtran}
\bibliography{ref}



%








\end{document}